\documentclass[aps,showpacs,floatfix,twocolumn,byrevtex,superscriptaddress]{revtex4-1}
\RequirePackage{lineno}
\usepackage{amsmath}
\usepackage{amssymb}
\usepackage{amstext}
\usepackage{amsopn}
\usepackage{amsfonts}
\usepackage{amsxtra}
\usepackage[english]{babel}
\usepackage{amsmath}
\usepackage{graphicx}
\usepackage{float}
\usepackage{bm}
\usepackage{multirow}
\usepackage{dcolumn}
\usepackage{inputenc}
\usepackage{hyperref}
\usepackage{CJK}
\usepackage{color}
\begin{document}

\title{Optical fingerprints of the electronic band reconstruction in van der Waals magnetic materials}

\author{M. Corasaniti$^{\dag}$}
\author{R. Yang$^{\dag}$}
\affiliation{Laboratorium f\"ur Festk\"orperphysik, ETH - Z\"urich, 8093 Z\"urich, Switzerland}

\author{Y. Liu$^{\ddag}$}
\affiliation{Condensed Matter Physics and Materials Science Department, Brookhaven National Laboratory, Upton NY 11973, USA}

\author{C. Petrovic}
\affiliation{Condensed Matter Physics and Materials Science Department, Brookhaven National Laboratory, Upton NY 11973, USA}

\author{L. Degiorgi$^\ast$}
\affiliation{Laboratorium f\"ur Festk\"orperphysik, ETH - Z\"urich, 8093 Z\"urich, Switzerland}

\date{\today}

\begin{abstract}
We report a broadband study of the charge dynamics in the van der Waals (vdW) magnetic materials 2H-$M_x$TaS$_2$ ($M$ = Mn and Co), which span the onset of both long-range antiferromagnetic (AFM) and ferromagnetic (FM) order, depending on the intercalation $M$ and its concentration $x$. We discover a spectral weight ($SW$) shift from high to low energy scales for FM compositions, while reversely $SW$ is removed from low towards high spectral energies for AFM compounds. This maps the related reconstruction of the electronic band structure along the crossover from the FM to AFM order, which restores an occupation balance in the density of states between spin majority and minority bands of the intercalated 3$d$ elements.
\end{abstract}
\maketitle

Understanding the physical mechanism as well as functionalities of van der Waals (vdW) heterostructures and electronic/spintronic devices is at present a central topic of the ongoing solid state physics research activities \cite{Zhang2019}. In this context, Fe$_3$GeTe$_2$ and MnSe$_2$ are prototype examples \cite{Deng2018,O'Hara2018} of two-dimensional vdW magnets. The discovery of such a long-range magnetism lately boosted the several decades long investigation of the (mostly non-magnetic) vdW transition-metal dichalcogenides (TMDCs) with appropriate magnetic intercalations. This enlarges the fundamental studies of low-dimensional magnetism and provides a platform for questing the nature of the critical behaviour of the spin interactions, ranging from Heisenberg and XY to Ising type \cite{Zhang2019}.

The TMDC materials of general formula TX$_2$ (e.g., with T = Nb, Ta and X = Se, S) consist of layered-like sandwiches held together by relatively weak forces across the vdW gaps, within which intercalation may occur. The latter is notably accompanied by charge transfer between the intercalated species and the host layer but is not supposed to change the local bonding within the sandwich layers, so that the electronic properties of the host TMDC materials could be treated within the rigid-band model and consequently be fine-tuned depending on the effective $d$-bands filling. Such a controllable degree of band filling seems to be a feature quite unique to low-dimensional structures \cite{Parkin1980,Friend1987}.

In this work, we focus our attention on the 2H-$M_x$TaS$_2$ compounds with $M$ = Co and Mn. While the possibility to intercalate TMDCs with magnetic 3$d$ elements $M$, exhibiting diverse magnetic properties, is known since the eighties, their physical properties have been only addressed recently \cite{Liu2021c,Liu2021,Liu2022}. Any intercalation with atoms, that tend to preserve local moment when embedded in a metallic host, induces long range magnetic order \cite{Zhang2021,Hatanaka2022}. This is different from the simple charge transfer that occurs upon intercalation of non-magnetic atom such as Pd or Li, and which commonly perturbs the charge-density-wave (CDW) and enhances superconductivity \cite{Bhoi2016,Liu2021b}. The pristine 2H-TaS$_2$ compound is in fact a familiar CDW material \cite{Harper1977} and the peculiarities with respect to its broken-symmetry ground state are amply documented in the literature. The (magnetic) Co and Mn intercalation (Fig. S1 in Supplemental Material (SM) \cite{SM}), besides removing the CDW transition, expands the vdW gap of the 2H-$M_x$TaS$_2$ crystals along the $c$-axis (i.e., orthogonal to the layers) and should result in an electronic doping via hybridization with atoms around the vdW gap \cite{Liu2022}. This induces a ferromagnetic (FM) state with an easy-plane anisotropy in 2H-Mn$_x$TaS$_2$ \cite{Hinode1995,Onuki1986, Li2011,Shand2015,Zhang2018,Liu2021}. For Co-intercalation, ferromagnetism with strong uniaxial anisotropy in 2H-Co$_{0.22}$TaS$_2$ \cite{Liu2021c} and a three-dimensional antiferromagnetic (AFM) state in 2H-Co$_{0.34}$TaS$_2$ \cite{VanLaar1971} have been firmly established. The evolving FM to AFM transition promoted by the Co-intercalation and the related sign-change of the ordinary Hall coefficient hint to an electronic multi-band occurrence and its reconstruction based on the magnetic ground state \cite{Liu2022}. The temperature-dependent electrical resistivity ($\rho(T)$) displays an overall metallic behaviour for all samples (Fig. S2 in SM \cite{SM}). A clear kink in $\rho(T)$, a weak anomaly in thermal conductivity, as well as a slope change in thermopower were yet observed at the magnetic transitions for 2H-Mn$_{0.28}$TaS$_2$ ($T_C \sim$ 82 K) and 2H-Co$_{0.34}$TaS$_2$ ($T_N \sim$  36 K), albeit weaker for crystals with lower concentration $x$ \cite{Liu2022}.

From a spectroscopic point of view, these materials are still not comprehensively scrutinised. Here, we investigate the temperature ($T$) dependence of the (in-plane) absorption spectrum \cite{Dressel2002} over a broad spectral range of both Co-and Mn-intercalated 2H-$M_x$TaS$_2$. 2H-Co$_x$TaS$_2$ harbour a distinct and reverse spectral weight ($SW$) reshuffling with regard to the FM and AFM transition at $T_C$ and $T_N$, respectively. The overall optical response of the FM Co-intercalated materials (i.e., $x \le$ 0.22) copies with the data collected on the Mn compositions, which all refer to a FM state. We supply arguments, which favour a reconstruction of the electronic band structure upon crossing over from the FM to AFM order with increasing Co-concentration, bearing testimony to the progressive lifting of the occupation imbalance in the density of states between spin majority and minority bands of the intercalated 3$d$ elements.

\begin{figure*}[!htb]
\center
\includegraphics[width=17.0cm]{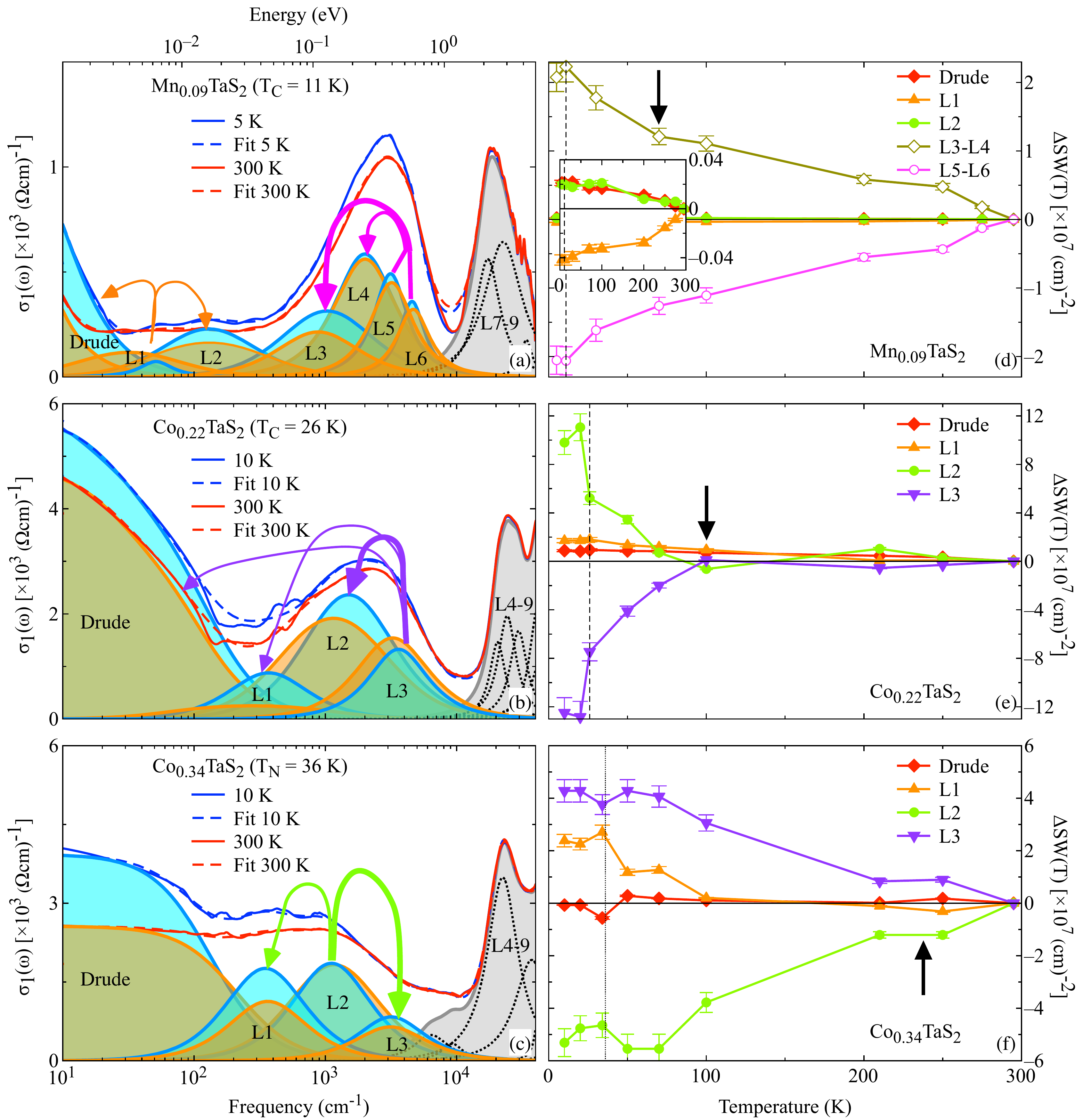}
\caption{(a-c) In-plane $\sigma_1(\omega)$ below 4$\times$10$^4$ cm$^{-1}$ (1 eV = 8.06548$\times$10$^3$ cm$^{-1}$, please note the logarithmic energy scale) at 5 or 10 and 300 K together with their respective total Drude-Lorentz fit (thick dashed line) after Eq. S1 in SM \cite{SM}, and (d-f) $T$ dependence of the $SW$ relative variation with respect to 300 K, i.e., $\Delta SW(T) = SW(T) - SW$(300 K) for selected fit components (see legend in panels (a-c)) of 2H-$M_x$TaS$_2$ ($M$ = Mn and Co): (a,d) $x$ = 0.09 (FM) Mn-concentration, and (b,e) $x$ = 0.22 (FM) as well as (c,f) $x$ = 0.34 (AFM) Co-concentration. Panels (a-c) explicitly show all fit components: the total Drude and Lorentz (Li, i = 1 to 9) HOs \cite{SM}. The coloured shaded areas emphasise $SW$ encountered by each component (reddish and blueish colours refer to 300 and 5 or 10 K, respectively, while the grey shaded area corresponds to $SW$ being $T$-independent). The rounded arrows in panels (a-c) highlight the direction in energy of the $SW$ reshuffling upon lowering $T$, which is stronger with thicker arrows. The inset in panel (d) is a blow-up of $\Delta SW$ for the total Drude term and HOs L1 and L2. The vertical dashed and dotted lines in panels (d-f) mark $T_C$ and $T_N$ (see Table I in SM \cite{SM}), respectively. The error bars in $\Delta SW(T)$ correspond to the direct propagation of the error in the HOs strength, estimated numerically within the non-linear least-squares fit technique. The vertical black arrows in panels (d-f) indicate $T^*$, as the onset of the faster $\Delta SW(T)$ variation upon lowering $T$. Additional data with similar analysis are available in Fig. S7 in SM \cite{SM}.
} 
\label{opt_cond_SW}
\end{figure*}

We launch first the survey about the $T$ dependence of the real part ($\sigma_1(\omega)$) of the optical conductivity, shown in Figs. \ref{opt_cond_SW}(a-c) for three selected Mn- and Co-concentration in the energy interval spanning the far- (FIR), mid- (MIR) and near- (NIR) infrared up to the visible spectral ranges at 5 or 10 and 300 K. We refer to SM in Ref.~\onlinecite{SM} for further details and additional data for other concentrations. The selected compositions are particularly pertinent, since they encompass both the FM and AFM magnetic phase transitions for the Co-intercalation and address a representative Mn compound towards its FM one (Table I in SM \cite{SM}). There is a metallic Drude-like component, which gets narrow as well as robust upon lowering $T$ (i.e., it generally gains $SW$, see below). It merges into a broad and $T$-dependent MIR absorption between 1000 and 3000 cm$^{-1}$ \cite{SM}.

In order to focus the discussion on the impact of the magnetic phase transition on the electronic properties, we propose their phenomenological Drude-Lorentz fit \cite{SM}, which is singled out for the spectra at 5 or 10 and 300 K in Figs. \ref{opt_cond_SW}(a-c). The chosen layout of the collected data allows to emphasise the $SW$ redistribution and its evolution as a function of $T$ among the Lorentz (Li, i = 1 to 9) harmonic oscillators (HO). In general, $SW$ of the optical conductivity corresponds to its integral $SW(T) = \frac{Z_0}{\pi^2}\int_{\omega_1}^{\omega_2}\sigma_1(\omega'; T)d\omega'$, expressed in units of cm$^{-2}$ ($Z_0$ = 376.73 $\Omega$, being the impedance of free space) \cite{Dressel2002}. $\omega_i$ ($i$ = 1 and 2) define the energy interval, relevant for the $SW$ estimation. Ahead, we alternatively propose to identify specific energy intervals via the phenomenological fit components, for which the related $SW$ corresponds to the square of the (Drude) plasma frequency or of the (Lorentz) HO strength (i.e., $\omega_{p,Di}^2$ or $\Omega_j^2$ in Eq. (S1) in SM \cite{SM}). As elaborated in SM \cite{SM}, the metallic part of $\sigma_1(\omega)$ necessitates of two Drude terms for the Co-intercalated materials, thus spotting the multiband nature of their electronic structure, while the Mn-intercalated compositions feature a single Drude component. The resulting global Drude $SW$ (i.e., $SW = \sum_i \omega_{p,Di}^2$, $i$ ranging over the number of contemplated Drude terms) will be anyhow at the centre of our attention.

The main findings of our work are summarised in Fig. \ref{opt_cond_SW}(d-f). First, in all compounds the $T$ dependence of the plasma frequencies (Fig. S8 in SM \cite{SM}) is such that the total Drude $SW$ either barely changes or moderately increases upon lowering $T$. For the Co-compositions, this also pairs with an additional $SW$ accumulating into the high frequency tail of the purely metallic response (represented by HO L1, Figs. \ref{opt_cond_SW}(b-c) and S7(b) in SM \cite{SM}). Such an enhancement results from a shift of $SW$ from higher energy scales. For the Mn-compounds, the Drude tail is given by the combination of HO L1 and L2 (Fig. \ref{opt_cond_SW}(a) and S7(a) in SM \cite{SM}), which equally suffer a $SW$ reordering among them and in favour of the Drude term. The $SW$ reshuffling affecting the Drude term and eventually its high-energy tail turns out though to be rather residual, compared to the $SW$ redistribution at higher energies (as e.g. emphasised by the inset in Fig. \ref{opt_cond_SW}(d) as well as in Fig. S7(c) in SM \cite{SM}). Second and even more relevant for our discussion, there is an opposite and strong redistribution of $SW$ between HO L2 and L3 depending on whether a FM or AFM transition (Figs. \ref{opt_cond_SW}(b-c)) takes place for the Co-intercalated compounds. For the Co-concentration $x$ = 0.22, HO L3 losses $SW$, which merges almost totally in HO L2 upon crossing $T_C$ (Fig. \ref{opt_cond_SW}(e)). The trend in the $SW$ removal and reallocation across the FM transition as observed in the Co-intercalated composition is similarly confirmed by the observations in the Mn-intercalated ones (e.g., for $x$ = 0.09 and 0.19 in Fig. \ref{opt_cond_SW}(d) and Fig. S7(c) in SM \cite{SM}, respectively). On the contrary, for the AFM Co-concentration $x$ = 0.34 we principally encounter a progressive and gradual shift of $SW$ from HO L2 towards HO L3 upon approaching $T_N$ from high $T$ (Fig. \ref{opt_cond_SW}(f)). An equivalent incidence is observed for the $x$ = 0.26 Co-intercalation (Fig. S7(d) in SM \cite{SM}), so that the $SW$ reshuffling for the AFM transition is fully exploited and leans towards a constant behaviour pattern below $T_N$. Summarising, we overall discern a mostly incremental $SW$ redistribution at FIR and MIR-NIR energy intervals upon approaching the magnetic phase transition from high $T$, which then tends to saturate into the magnetic state (i.e., at $T < T_N$ or $T_C$). Additionally, a change of slope in $\Delta SW(T)$ is observed in all compositions at $T$ ranging between $\sim$ 100 and 250 K. This is noted by $T^*$ (black arrows in Figs. \ref{opt_cond_SW}(d-f) and S7(c-d) in SM \cite{SM}). The resulting kink in $\Delta SW(T)$ is smooth in the Mn-intercalated materials and rather abrupt and sudden in the Co-ones (at least for $x$ = 0.22 (Fig. \ref{opt_cond_SW}(e)) and 0.26 (Fig. S7(d) in SM \cite{SM}) and somehow stepwise for $x$ = 0.34 (Fig. \ref{opt_cond_SW}(f)). These observations fairly agree with similar findings (as kink or upturn around $T^*$) in the measured $T$ dependence of the in-plane thermopower and total thermal conductivity as well as Hall resistivity \cite{Liu2022}. The origin of these peculiarities and their link to the advanced multiband nature of these materials need to be better understood, as offered here from the optical perspective.

The $SW$ distribution and its evolution upon crossing $T_C$ or $T_N$ seems to be a common property for both Co or Mn intercalations and is exclusively driven by the targeted, final magnetic state (i.e., independent of the element choice). Moreover, the encountered shift of $SW$ occurs at FIR-MIR energy scales up to the NIR spectral range, while at visible and ultra-violet frequencies $SW$ is constant at any $T$ (grey shaded areas in Figs. \ref{opt_cond_SW}(a-c) and Figs. S7(a-b) in SM \cite{SM}). This also means that the full recovery of $SW$ is achieved at about 1 eV. Nonetheless, the opposite $SW$ allocation discovered upon lowering $T$ through either a FM or AFM transition (see rounded arrows in Figs. \ref{opt_cond_SW}(a-c) and Figs. S7(a-b) in SM \cite{SM}) calls for a yet different reconstruction of the related electronic band structure, upon which we wish to argue for the rest of our paper.

\begin{figure}[tb]
\centerline{
\includegraphics[width=1\columnwidth]{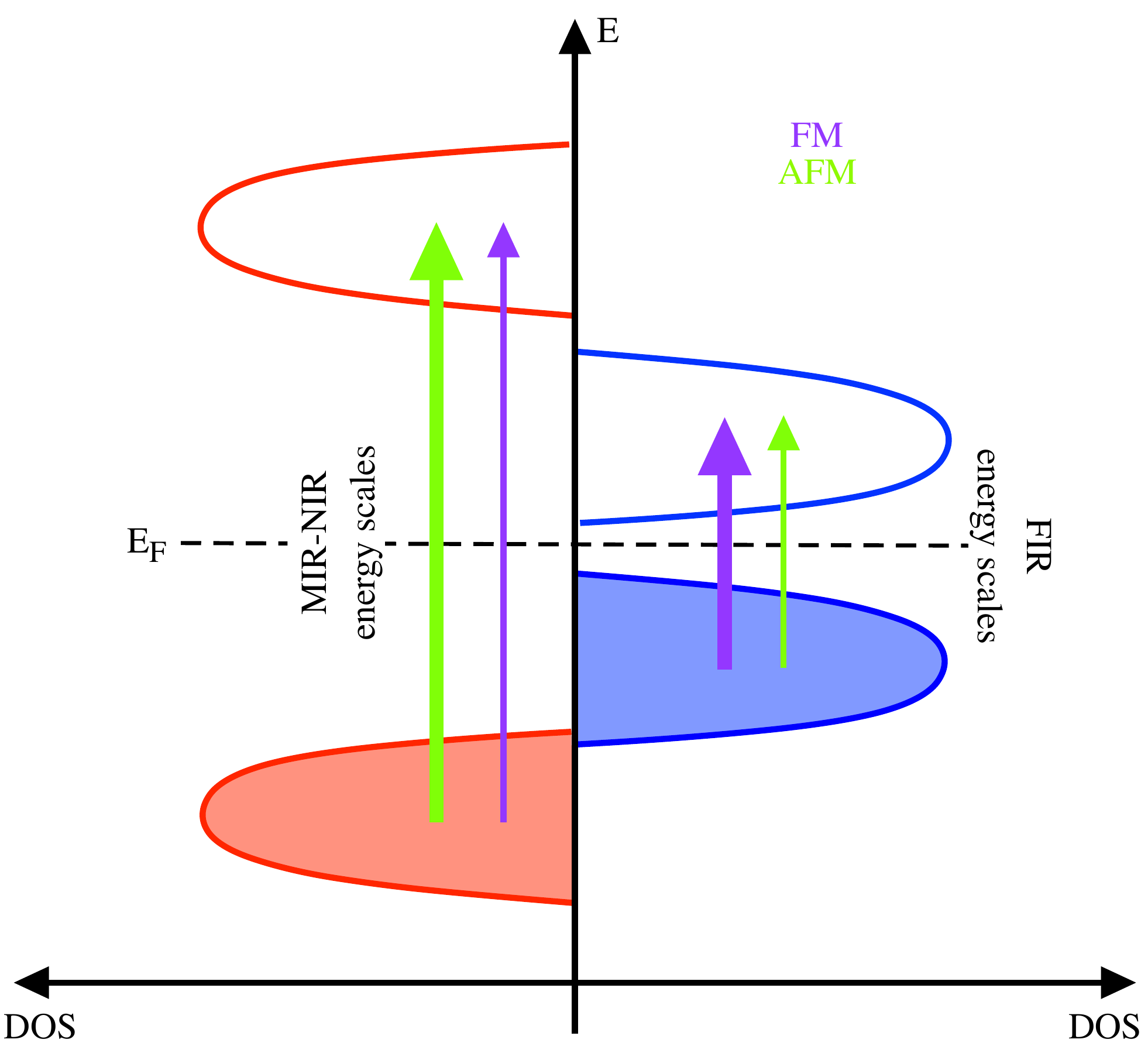}}
\caption{Proposal for DOS particularly emphasising the interband transitions grouping around the characteristic FIR energy scales of 0.05-0.2 eV (i.e., between the blue bands) and MIR-NIR ones of 0.3-0.5 eV (i.e., between the red bands). The trend in the related transition probability (i.e., $SW$ redistribution of the proposed interband excitations) is indicated by the arrows thickness (violet vertical arrows for the FM state and green vertical arrows for the AFM state). The colour code of the arrows is in accord with the convention used for the $SW$ reshuffling in the AFM and FM state of Figs. \ref{opt_cond_SW}(a-c) and Figs. S7(a-b) in SM \cite{SM}.
} 
\label{scheme_band_structure}
\end{figure}

Figure \ref{scheme_band_structure} schematically depicts the density-of-states (DOS), factual for the alleged progression of the interband transition probability upon lowering $T$ across the FM and AFM transitions. The charge dynamics and the evolution of its stored $SW$ convey the presence of interband transitions grouping within two distinct spectral intervals, i.e., at FIR resonance energies between 50 and 200 meV (i.e., involving the blue bands in Fig. \ref{scheme_band_structure}) as well as at MIR-NIR ones between 300 and 500 meV (i.e., involving the red bands in Fig. \ref{scheme_band_structure}). These ranges are described by the combination of HOs L3-L4 and L5-L6 for the Mn-intercalated material and L2 and L3 for the Co-intercalated compositions (Figs. \ref{opt_cond_SW}(a-c)), respectively. The thickness of the coloured (vertical) arrows in Fig. \ref{scheme_band_structure} mimics the strength (i.e., $SW$) of those two possible interband transitions, which alike discriminate between the two magnetic states. Below $T_N$, the (convoluted) MIR-NIR transition is stronger than the lower FIR one, while the opposite seems to apply below $T_C$.

The characteristic ingredients pertinent to the reconstruction of the electronic band structure, depicted in Fig. \ref{scheme_band_structure} and then driving the $SW$ redistribution observed in $\sigma_1(\omega)$ (Fig. \ref{opt_cond_SW} and Fig. S7 in SM \cite{SM}), embrace several findings and achieved knowledge on related, sister materials. First of all and from a general perspective, the FM state in both Mn- and Co-intercalated 2H-TaS$_2$ has been tentatively reconciled within a Ruderman-Kittel-Kasuya-Yosida interaction scenario \cite{Liu2021c,Liu2021,Liu2022}, in which the local spins of intercalated Mn and Co ions align ferromagnetically through the itinerant Ta 5$d$ electrons, as initially proposed for the related Co-intercalated 2H-NbS$_2$ material \cite{Barisic2011}. Further, it is speculated that Co atoms tend to hybridise more strongly with the electronic states associated with covalently bonded structural subunit (i.e., Ta and S) when intercalated in the vdW gap. By enlarging the Co concentration, this leads to a suppression of the spontaneous magnetic moment and to a stronger tendency for AFM exchange coupling parameters \cite{Liu2022,Polesya2019}. 

Along this line of thoughts, the FM to AFM crossover, hither studied upon changing the intercalation and contingently the concentration of the intercalated Co atom, can be also achieved after two alternative ways: either by differentiating the element-intercalation at given concentration or by applying pressure on a selected intercalated compound. It is experimentally known that the structurally equivalent 2H-NbS$_2$ compound \cite{Structure}, intercalated with 3$d$ elements Cr, Mn and Fe for the concentration 1/3, exhibits a variety of magnetic states, (roughly) classified as FM for the Cr and similarly Mn materials \cite{Hulliger1970,Miyadai1983,VanLaar1971,Kousaka2009} and as AFM for the Fe composition \cite{VanLaar1971,Gorochov1981,Yamamura2004}. The first-principles electronic band structure calculations that have been performed using the fully relativistic Korringa-Kohn-Rostoker Green function method \cite{Mankovsky2016} have pointed out the complexity of their magnetic ordering. For Cr$_{1/3}$NbS$_2$ and Mn$_{1/3}$NbS$_2$, the in-plane magnetocrystalline anisotropy and Dzyaloshinskii-Moriya interactions give rise to a helimagnetic structure along the $c$-axis, following the experimental observations \cite{Hulliger1970,Miyadai1983}. On the other hand, the negative exchange interactions in the Fe$_{1/3}$NbS$_2$ compound result in a noncollinear frustrated magnetic structure if the magnetocrystalline anisotropy is not taken into account. However, a strong magnetocrystalline anisotropy along the $c$-axis does lead to a magnetic state referred to as an ordering of the third kind, which was indeed determined experimentally \cite{VanLaar1971,Gorochov1981,Yamamura2004}. This may be pinned down to the diverse DOS with respect to the magnetic state in these series of compounds. For all compounds, DOS is rather large for the majority-spin states at the Fermi energy ($E_F$). In the case of minority-spin states of Cr$_{1/3}$NbS$_2$ and Mn$_{1/3}$NbS$_2$, one can however observe a pseudogap between the occupied and unoccupied states \cite{Polesya2019,Mankovsky2016}. Such an imbalance then leads to the FM ground state. In the case of Fe$_{1/3}$NbS$_2$, $E_F$ is located at the DOS maximum corresponding to the Fe minority-spin $d$ states. DOS is thus finite for both spin directions, which facilitates an AFM ground state \cite{Mankovsky2016}. This latter trend is reflected in the isotropic exchange coupling parameter for the three compositions, which is predominantly positive for the Cr and Mn compounds (i.e., promoting FM) but negative (i.e., leading to AFM) for Fe-Fe interactions at short distance \cite{Mankovsky2016}.

In turn, one can induce the crossover from the FM to AFM state in Mn$_{1/4}$NbS$_2$ upon applying pressure \cite{Polesya2020}. This possibility is instrumental, in order to better justify and support our schematic proposal in Fig. \ref{scheme_band_structure}. Focusing the attention on the spin-resolved DOS on Mn sites, which is foremost contributed by the $d$-orbitals, the majority-spin states are almost occupied around both the $\Gamma$ and $K$ points in the Brillouin zone at ambient pressure, while the minority-spin states, essentially around the $K$ point of the Brillouin zone, are unoccupied. This favors FM and further implies a dominating hole-type character of the electric carriers and a positive slope of the Hall resistivity \cite{Polesya2020}. Interband transitions within the FIR energy interval (so roughly peaked at 100 meV) are foreseen from the electronic band structure at both $\Gamma$ and $K$ points of the Brillouin zone \cite{Polesya2020} and should play the most prominent role in $\sigma_1(\omega)$. Figure \ref{scheme_band_structure} catches a glimpse of such a trim for our FM compositions (Figs. \ref{opt_cond_SW}(a-b) and S7(a) in SM \cite{SM}). Conversely, the pressure increase results in the broadening of the energy bands, which premises an occupation of the bottom of the minority-spin states (having mainly $d_{x^2-y^2}$ and $d_{xy}$ character) and a draining of the top of minority-spin states (primarily of $d_{xz}$and $d_{yz}$ character) at the $K$ point of the Brillouin zone in Mn$_{1/4}$NbS$_2$ \cite{Polesya2020}. This is accompanied by a decrease of the exchange splitting of the majority- and minority-spin $d$-states of Mn, which potentially turns negative for the first neighbour interaction, as the prerequisite for an AFM alignment of the magnetic moments in the absence of any other interactions. In fact, such a setting is translated into an electron-like character of the ordinary Hall effect \cite{Polesya2020}. By inspecting the resulting electronic band structure in the AFM state \cite{Polesya2020}, we recognise the most cogent consequence for the excitation spectrum: namely, low energy FIR interband transitions are less probable, while the probability for high energy ones at MIR and NIR frequencies (i.e., settled around 400 meV) increases sensitively, as evinced in $\sigma_1(\omega)$ (Figs. \ref{opt_cond_SW}(c) and S7(b) in SM \cite{SM}) and as also sketched in Fig. \ref{scheme_band_structure} for our case. Therefore, we claim that the band reconstruction in Mn$_{1/4}$NbS$_2$ upon applying pressure is implementable to 2H-Co$_x$TaS$_2$ with respect to the FM-AFM crossover as a function of $x$, as backed up by the same sign-change of the ordinary Hall coefficient upon varying the magnetic ground state \cite{Liu2022,Polesya2020}.

In conclusion, the charge dynamics of 2H-$M_x$TaS$_2$ ($M$ = Mn and Co) allows to determine a distinct $SW$ redistribution and to shed light on the relevant energy scales shaping the reconstruction of the electronic band structure upon crossing over from the FM to AFM order with varying intercalation (i.e., element and/or its concentration). Our spectroscopic findings seem to be consistent with dedicated first-principles calculations upon tuning element-intercalation, pressure and/or magnetic field on similar intercalated materials. The applicability of the proposed comparison is validated by the fact, that the quoted calculations do tackle the impact of (similar) magnetic transitions on the electronic band structure of equivalent materials, seemingly reflected in the charge dynamics of our compositions, too. 

Summing up, the significance of this work consists in the first instance in the systematic investigation of 2H-TaS$_2$ with different intercalations and across distinct magnetic transitions, hence widening out a previous, more restricted attempt \cite{Hu2008} and possibly challenging the so far broadly accepted implementation of the rigid-band model. Moreover, since the continuous change of the dominant magnetic exchange controls the FM to AFM switching and being its impact mapped onto the easily accessible FIR-MIR-NIR spectral range, one may exploit our straightforward (broadband) experimental tool also as a function of alternative tuneable variables than chemical-intercalation, which would thoroughly flash on the overall consistency of the emerging physical picture. Finally, the present work demonstrates a feasible route towards understanding magnetism in low dimensions and may help in revealing robust properties, relevant for the development of low-power spin-logic circuits from layered materials \cite{Liu2021c}. The possibility of integration of FM is generally of interest for spintronics, so that novel fabrication of low-dimensional heterostructures might be envisaged and motivated, as well; the few-layer graphene/2H-TaS$_2$ heterostructures with robust spin-helical state \cite{Li2020} is already a promising development and the recent discovery of one-dimensional vdW (yet non magnetic) heterostructures \cite{Xiang2020} and multi-walled 2H-TaS$_2$ nanotubes \cite{Li2010} may open new avenues. Since exotic phenomena such as skyrmions and magnetic solitons \cite{Zhang2022,Zheng2021} were recently discovered in 2H-TaS$_2$-based vdW magnets, our results may therefore help to efficiently tune their properties for further applications.\\

\section*{Acknowledgements}
Work at Brookhaven National Laboratory was supported by the U.S. Department of Energy, Office of Basic Energy Science, Division of Materials Science and Engineering, under Contract No. DE-SC0012704 (materials synthesis). \\

$^{\dag}$ Authors M.C. and R.Y. contributed equally to the work.\\

$^{\ddag}$ Present address: Los Alamos National Laboratory, Los Alamos, New Mexico 87545\\

$^\ast$ Correspondence and requests for materials should be addressed to: 
L. Degiorgi, Laboratorium f\"ur Festk\"orperphysik, ETH - Z\"urich, 8093 Z\"urich, Switzerland; 
email: degiorgi@solid.phys.ethz.ch.


\begin{thebibliography}{40}%
\makeatletter
\providecommand \@ifxundefined [1]{%
 \@ifx{#1\undefined}
}%
\providecommand \@ifnum [1]{%
 \ifnum #1\expandafter \@firstoftwo
 \else \expandafter \@secondoftwo
 \fi
}%
\providecommand \@ifx [1]{%
 \ifx #1\expandafter \@firstoftwo
 \else \expandafter \@secondoftwo
 \fi
}%
\providecommand \natexlab [1]{#1}%
\providecommand \enquote  [1]{``#1''}%
\providecommand \bibnamefont  [1]{#1}%
\providecommand \bibfnamefont [1]{#1}%
\providecommand \citenamefont [1]{#1}%
\providecommand \href@noop [0]{\@secondoftwo}%
\providecommand \href [0]{\begingroup \@sanitize@url \@href}%
\providecommand \@href[1]{\@@startlink{#1}\@@href}%
\providecommand \@@href[1]{\endgroup#1\@@endlink}%
\providecommand \@sanitize@url [0]{\catcode `\\12\catcode `\$12\catcode
  `\&12\catcode `\#12\catcode `\^12\catcode `\_12\catcode `\%12\relax}%
\providecommand \@@startlink[1]{}%
\providecommand \@@endlink[0]{}%
\providecommand \url  [0]{\begingroup\@sanitize@url \@url }%
\providecommand \@url [1]{\endgroup\@href {#1}{\urlprefix }}%
\providecommand \urlprefix  [0]{URL }%
\providecommand \Eprint [0]{\href }%
\providecommand \doibase [0]{http://dx.doi.org/}%
\providecommand \selectlanguage [0]{\@gobble}%
\providecommand \bibinfo  [0]{\@secondoftwo}%
\providecommand \bibfield  [0]{\@secondoftwo}%
\providecommand \translation [1]{[#1]}%
\providecommand \BibitemOpen [0]{}%
\providecommand \bibitemStop [0]{}%
\providecommand \bibitemNoStop [0]{.\EOS\space}%
\providecommand \EOS [0]{\spacefactor3000\relax}%
\providecommand \BibitemShut  [1]{\csname bibitem#1\endcsname}%
\let\auto@bib@innerbib\@empty
\bibitem [{\citenamefont {Zhang}\ \emph {et~al.}(2019)\citenamefont {Zhang},
  \citenamefont {Wong}, \citenamefont {Zhu},\ and\ \citenamefont
  {Wee}}]{Zhang2019}%
  \BibitemOpen
  \bibfield  {author} {\bibinfo {author} {\bibfnamefont {W.}~\bibnamefont
  {Zhang}}, \bibinfo {author} {\bibfnamefont {P.~K.~J.}\ \bibnamefont {Wong}},
  \bibinfo {author} {\bibfnamefont {R.}~\bibnamefont {Zhu}}, \ and\ \bibinfo
  {author} {\bibfnamefont {A.~T.~S.}\ \bibnamefont {Wee}},\ }\href {\doibase
  https://doi.org/10.1002/inf2.12048} {\bibfield  {journal} {\bibinfo
  {journal} {InfoMat}\ }\textbf {\bibinfo {volume} {1}},\ \bibinfo {pages}
  {479} (\bibinfo {year} {2019})}\BibitemShut {NoStop}%
\bibitem [{\citenamefont {Deng}\ \emph {et~al.}(2018)\citenamefont {Deng},
  \citenamefont {Yu}, \citenamefont {Song}, \citenamefont {Zhang},
  \citenamefont {Wang}, \citenamefont {Sun}, \citenamefont {Yi}, \citenamefont
  {Wu}, \citenamefont {Wu}, \citenamefont {Zhu}, \citenamefont {Wang},
  \citenamefont {Chen},\ and\ \citenamefont {Zhang}}]{Deng2018}%
  \BibitemOpen
  \bibfield  {author} {\bibinfo {author} {\bibfnamefont {Y.}~\bibnamefont
  {Deng}}, \bibinfo {author} {\bibfnamefont {Y.}~\bibnamefont {Yu}}, \bibinfo
  {author} {\bibfnamefont {Y.}~\bibnamefont {Song}}, \bibinfo {author}
  {\bibfnamefont {J.}~\bibnamefont {Zhang}}, \bibinfo {author} {\bibfnamefont
  {N.~Z.}\ \bibnamefont {Wang}}, \bibinfo {author} {\bibfnamefont
  {Z.}~\bibnamefont {Sun}}, \bibinfo {author} {\bibfnamefont {Y.}~\bibnamefont
  {Yi}}, \bibinfo {author} {\bibfnamefont {Y.~Z.}\ \bibnamefont {Wu}}, \bibinfo
  {author} {\bibfnamefont {S.}~\bibnamefont {Wu}}, \bibinfo {author}
  {\bibfnamefont {J.}~\bibnamefont {Zhu}}, \bibinfo {author} {\bibfnamefont
  {J.}~\bibnamefont {Wang}}, \bibinfo {author} {\bibfnamefont {X.~H.}\
  \bibnamefont {Chen}}, \ and\ \bibinfo {author} {\bibfnamefont
  {Y.}~\bibnamefont {Zhang}},\ }\href {\doibase 10.1038/s41586-018-0626-9}
  {\bibfield  {journal} {\bibinfo  {journal} {Nature}\ }\textbf {\bibinfo
  {volume} {563}},\ \bibinfo {pages} {94} (\bibinfo {year} {2018})}\BibitemShut
  {NoStop}%
\bibitem [{\citenamefont {O'Hara}\ \emph {et~al.}(2018)\citenamefont {O'Hara},
  \citenamefont {Zhu}, \citenamefont {Trout}, \citenamefont {Ahmed},
  \citenamefont {Luo}, \citenamefont {Lee}, \citenamefont {Brenner},
  \citenamefont {Rajan}, \citenamefont {Gupta}, \citenamefont {McComb},\ and\
  \citenamefont {Kawakami}}]{O'Hara2018}%
  \BibitemOpen
  \bibfield  {author} {\bibinfo {author} {\bibfnamefont {D.~J.}\ \bibnamefont
  {O'Hara}}, \bibinfo {author} {\bibfnamefont {T.}~\bibnamefont {Zhu}},
  \bibinfo {author} {\bibfnamefont {A.~H.}\ \bibnamefont {Trout}}, \bibinfo
  {author} {\bibfnamefont {A.~S.}\ \bibnamefont {Ahmed}}, \bibinfo {author}
  {\bibfnamefont {Y.~K.}\ \bibnamefont {Luo}}, \bibinfo {author} {\bibfnamefont
  {C.~H.}\ \bibnamefont {Lee}}, \bibinfo {author} {\bibfnamefont {M.~R.}\
  \bibnamefont {Brenner}}, \bibinfo {author} {\bibfnamefont {S.}~\bibnamefont
  {Rajan}}, \bibinfo {author} {\bibfnamefont {J.~A.}\ \bibnamefont {Gupta}},
  \bibinfo {author} {\bibfnamefont {D.~W.}\ \bibnamefont {McComb}}, \ and\
  \bibinfo {author} {\bibfnamefont {R.~K.}\ \bibnamefont {Kawakami}},\ }\href
  {\doibase 10.1021/acs.nanolett.8b00683} {\bibfield  {journal} {\bibinfo
  {journal} {Nano Letters}\ }\textbf {\bibinfo {volume} {18}},\ \bibinfo
  {pages} {3125} (\bibinfo {year} {2018})}\BibitemShut {NoStop}%
\bibitem [{\citenamefont {Parkin}\ and\ \citenamefont
  {Friend}(1980)}]{Parkin1980}%
  \BibitemOpen
  \bibfield  {author} {\bibinfo {author} {\bibfnamefont {S.~S.~P.}\
  \bibnamefont {Parkin}}\ and\ \bibinfo {author} {\bibfnamefont {R.~H.}\
  \bibnamefont {Friend}},\ }\href {\doibase 10.1080/13642818008245370}
  {\bibfield  {journal} {\bibinfo  {journal} {Philosophical Magazine B}\
  }\textbf {\bibinfo {volume} {41}},\ \bibinfo {pages} {65} (\bibinfo {year}
  {1980})}\BibitemShut {NoStop}%
\bibitem [{\citenamefont {Friend}\ and\ \citenamefont
  {Yoffe}(1987)}]{Friend1987}%
  \BibitemOpen
  \bibfield  {author} {\bibinfo {author} {\bibfnamefont {R.}~\bibnamefont
  {Friend}}\ and\ \bibinfo {author} {\bibfnamefont {A.}~\bibnamefont {Yoffe}},\
  }\href {\doibase 10.1080/00018738700101951} {\bibfield  {journal} {\bibinfo
  {journal} {Advances in Physics}\ }\textbf {\bibinfo {volume} {36}},\ \bibinfo
  {pages} {1} (\bibinfo {year} {1987})}\BibitemShut {NoStop}%
\bibitem [{\citenamefont {Liu}\ \emph {et~al.}(2021{\natexlab{a}})\citenamefont
  {Liu}, \citenamefont {Hu}, \citenamefont {Stavitski}, \citenamefont
  {Attenkofer},\ and\ \citenamefont {Petrovic}}]{Liu2021c}%
  \BibitemOpen
  \bibfield  {author} {\bibinfo {author} {\bibfnamefont {Y.}~\bibnamefont
  {Liu}}, \bibinfo {author} {\bibfnamefont {Z.}~\bibnamefont {Hu}}, \bibinfo
  {author} {\bibfnamefont {E.}~\bibnamefont {Stavitski}}, \bibinfo {author}
  {\bibfnamefont {K.}~\bibnamefont {Attenkofer}}, \ and\ \bibinfo {author}
  {\bibfnamefont {C.}~\bibnamefont {Petrovic}},\ }\href {\doibase
  10.1103/PhysRevResearch.3.023181} {\bibfield  {journal} {\bibinfo  {journal}
  {Phys. Rev. Research}\ }\textbf {\bibinfo {volume} {3}},\ \bibinfo {pages}
  {023181} (\bibinfo {year} {2021}{\natexlab{a}})}\BibitemShut {NoStop}%
\bibitem [{\citenamefont {Liu}\ \emph {et~al.}(2021{\natexlab{b}})\citenamefont
  {Liu}, \citenamefont {Hu}, \citenamefont {Stavitski}, \citenamefont
  {Attenkofer},\ and\ \citenamefont {Petrovic}}]{Liu2021}%
  \BibitemOpen
  \bibfield  {author} {\bibinfo {author} {\bibfnamefont {Y.}~\bibnamefont
  {Liu}}, \bibinfo {author} {\bibfnamefont {Z.}~\bibnamefont {Hu}}, \bibinfo
  {author} {\bibfnamefont {E.}~\bibnamefont {Stavitski}}, \bibinfo {author}
  {\bibfnamefont {K.}~\bibnamefont {Attenkofer}}, \ and\ \bibinfo {author}
  {\bibfnamefont {C.}~\bibnamefont {Petrovic}},\ }\href {\doibase
  10.1103/PhysRevB.103.144432} {\bibfield  {journal} {\bibinfo  {journal}
  {Phys. Rev. B}\ }\textbf {\bibinfo {volume} {103}},\ \bibinfo {pages}
  {144432} (\bibinfo {year} {2021}{\natexlab{b}})}\BibitemShut {NoStop}%
\bibitem [{\citenamefont {Liu}\ \emph {et~al.}(2022)\citenamefont {Liu},
  \citenamefont {Hu}, \citenamefont {Tong}, \citenamefont {Bauer},\ and\
  \citenamefont {Petrovic}}]{Liu2022}%
  \BibitemOpen
  \bibfield  {author} {\bibinfo {author} {\bibfnamefont {Y.}~\bibnamefont
  {Liu}}, \bibinfo {author} {\bibfnamefont {Z.}~\bibnamefont {Hu}}, \bibinfo
  {author} {\bibfnamefont {X.}~\bibnamefont {Tong}}, \bibinfo {author}
  {\bibfnamefont {E.~D.}\ \bibnamefont {Bauer}}, \ and\ \bibinfo {author}
  {\bibfnamefont {C.}~\bibnamefont {Petrovic}},\ }\href {\doibase
  10.1103/PhysRevResearch.4.013048} {\bibfield  {journal} {\bibinfo  {journal}
  {Phys. Rev. Research}\ }\textbf {\bibinfo {volume} {4}},\ \bibinfo {pages}
  {013048} (\bibinfo {year} {2022})}\BibitemShut {NoStop}%
\bibitem [{\citenamefont {Zhang}\ \emph {et~al.}(2021)\citenamefont {Zhang},
  \citenamefont {Zhang}, \citenamefont {Liu}, \citenamefont {Zhang},
  \citenamefont {Yuan}, \citenamefont {Li}, \citenamefont {Wen}, \citenamefont
  {Jiang}, \citenamefont {Zhou}, \citenamefont {Lei}, \citenamefont {Zheng},
  \citenamefont {Song}, \citenamefont {Hou}, \citenamefont {Mi}, \citenamefont
  {Schwingenschl\"ogl}, \citenamefont {Manchon}, \citenamefont {Qiu},
  \citenamefont {Alshareef}, \citenamefont {Peng},\ and\ \citenamefont
  {Zhang}}]{Zhang2021}%
  \BibitemOpen
  \bibfield  {author} {\bibinfo {author} {\bibfnamefont {C.}~\bibnamefont
  {Zhang}}, \bibinfo {author} {\bibfnamefont {J.}~\bibnamefont {Zhang}},
  \bibinfo {author} {\bibfnamefont {C.}~\bibnamefont {Liu}}, \bibinfo {author}
  {\bibfnamefont {S.}~\bibnamefont {Zhang}}, \bibinfo {author} {\bibfnamefont
  {Y.}~\bibnamefont {Yuan}}, \bibinfo {author} {\bibfnamefont {P.}~\bibnamefont
  {Li}}, \bibinfo {author} {\bibfnamefont {Y.}~\bibnamefont {Wen}}, \bibinfo
  {author} {\bibfnamefont {Z.}~\bibnamefont {Jiang}}, \bibinfo {author}
  {\bibfnamefont {B.}~\bibnamefont {Zhou}}, \bibinfo {author} {\bibfnamefont
  {Y.}~\bibnamefont {Lei}}, \bibinfo {author} {\bibfnamefont {D.}~\bibnamefont
  {Zheng}}, \bibinfo {author} {\bibfnamefont {C.}~\bibnamefont {Song}},
  \bibinfo {author} {\bibfnamefont {Z.}~\bibnamefont {Hou}}, \bibinfo {author}
  {\bibfnamefont {W.}~\bibnamefont {Mi}}, \bibinfo {author} {\bibfnamefont
  {U.}~\bibnamefont {Schwingenschl\"ogl}}, \bibinfo {author} {\bibfnamefont
  {A.}~\bibnamefont {Manchon}}, \bibinfo {author} {\bibfnamefont {Z.~Q.}\
  \bibnamefont {Qiu}}, \bibinfo {author} {\bibfnamefont {H.~N.}\ \bibnamefont
  {Alshareef}}, \bibinfo {author} {\bibfnamefont {Y.}~\bibnamefont {Peng}}, \
  and\ \bibinfo {author} {\bibfnamefont {X.-X.}\ \bibnamefont {Zhang}},\ }\href
  {\doibase https://doi.org/10.1002/adma.202101131} {\bibfield  {journal}
  {\bibinfo  {journal} {Advanced Materials}\ }\textbf {\bibinfo {volume}
  {33}},\ \bibinfo {pages} {2101131} (\bibinfo {year} {2021})}\BibitemShut
  {NoStop}%
\bibitem [{\citenamefont {Hatanaka}\ \emph {et~al.}(2022)\citenamefont
  {Hatanaka}, \citenamefont {Nomoto},\ and\ \citenamefont
  {Arita}}]{Hatanaka2022}%
  \BibitemOpen
  \bibfield  {author} {\bibinfo {author} {\bibfnamefont {T.}~\bibnamefont
  {Hatanaka}}, \bibinfo {author} {\bibfnamefont {T.}~\bibnamefont {Nomoto}}, \
  and\ \bibinfo {author} {\bibfnamefont {R.}~\bibnamefont {Arita}},\ }\href
  {\doibase 10.48550/ARXIV.2210.07740} {\  (\bibinfo {year} {2022}),\
  10.48550/ARXIV.2210.07740}\BibitemShut {NoStop}%
\bibitem [{\citenamefont {Bhoi}\ \emph {et~al.}(2016)\citenamefont {Bhoi},
  \citenamefont {Khim}, \citenamefont {Nam}, \citenamefont {Lee}, \citenamefont
  {Kim}, \citenamefont {Jeon}, \citenamefont {Min}, \citenamefont {Park},\ and\
  \citenamefont {Kim}}]{Bhoi2016}%
  \BibitemOpen
  \bibfield  {author} {\bibinfo {author} {\bibfnamefont {D.}~\bibnamefont
  {Bhoi}}, \bibinfo {author} {\bibfnamefont {S.}~\bibnamefont {Khim}}, \bibinfo
  {author} {\bibfnamefont {W.}~\bibnamefont {Nam}}, \bibinfo {author}
  {\bibfnamefont {B.~S.}\ \bibnamefont {Lee}}, \bibinfo {author} {\bibfnamefont
  {C.}~\bibnamefont {Kim}}, \bibinfo {author} {\bibfnamefont {B.~G.}\
  \bibnamefont {Jeon}}, \bibinfo {author} {\bibfnamefont {B.~H.}\ \bibnamefont
  {Min}}, \bibinfo {author} {\bibfnamefont {S.}~\bibnamefont {Park}}, \ and\
  \bibinfo {author} {\bibfnamefont {K.~H.}\ \bibnamefont {Kim}},\ }\href
  {\doibase 10.1038/srep24068} {\bibfield  {journal} {\bibinfo  {journal}
  {Scientific Reports}\ }\textbf {\bibinfo {volume} {6}},\ \bibinfo {pages}
  {24068} (\bibinfo {year} {2016})}\BibitemShut {NoStop}%
\bibitem [{\citenamefont {Liu}\ \emph {et~al.}(2021{\natexlab{c}})\citenamefont
  {Liu}, \citenamefont {Huangfu}, \citenamefont {Zhang}, \citenamefont {Lin},\
  and\ \citenamefont {Schilling}}]{Liu2021b}%
  \BibitemOpen
  \bibfield  {author} {\bibinfo {author} {\bibfnamefont {H.}~\bibnamefont
  {Liu}}, \bibinfo {author} {\bibfnamefont {S.}~\bibnamefont {Huangfu}},
  \bibinfo {author} {\bibfnamefont {X.}~\bibnamefont {Zhang}}, \bibinfo
  {author} {\bibfnamefont {H.}~\bibnamefont {Lin}}, \ and\ \bibinfo {author}
  {\bibfnamefont {A.}~\bibnamefont {Schilling}},\ }\href {\doibase
  10.1103/PhysRevB.104.064511} {\bibfield  {journal} {\bibinfo  {journal}
  {Phys. Rev. B}\ }\textbf {\bibinfo {volume} {104}},\ \bibinfo {pages}
  {064511} (\bibinfo {year} {2021}{\natexlab{c}})}\BibitemShut {NoStop}%
\bibitem [{\citenamefont {Harper}\ \emph {et~al.}(1977)\citenamefont {Harper},
  \citenamefont {Geballe},\ and\ \citenamefont {DiSalvo}}]{Harper1977}%
  \BibitemOpen
  \bibfield  {author} {\bibinfo {author} {\bibfnamefont {J.~M.~E.}\
  \bibnamefont {Harper}}, \bibinfo {author} {\bibfnamefont {T.~H.}\
  \bibnamefont {Geballe}}, \ and\ \bibinfo {author} {\bibfnamefont {F.~J.}\
  \bibnamefont {DiSalvo}},\ }\href {\doibase 10.1103/PhysRevB.15.2943}
  {\bibfield  {journal} {\bibinfo  {journal} {Phys. Rev. B}\ }\textbf {\bibinfo
  {volume} {15}},\ \bibinfo {pages} {2943} (\bibinfo {year}
  {1977})}\BibitemShut {NoStop}%
\bibitem [{SM()}]{SM}%
  \BibitemOpen
  \href@noop {} {}\bibinfo {note} {See Supplemental Material at [URL will be
  inserted by publisher] for additional information on the samples growth and
  properties, experimental techniques and data analysis, which includes the
  additional Refs. \onlinecite{Hu2007,Maulana2021,Corasaniti2020}.}\BibitemShut
  {Stop}%
\bibitem [{\citenamefont {Hinode}\ \emph {et~al.}(1995)\citenamefont {Hinode},
  \citenamefont {Ohtani},\ and\ \citenamefont {Wakihara}}]{Hinode1995}%
  \BibitemOpen
  \bibfield  {author} {\bibinfo {author} {\bibfnamefont {H.}~\bibnamefont
  {Hinode}}, \bibinfo {author} {\bibfnamefont {T.}~\bibnamefont {Ohtani}}, \
  and\ \bibinfo {author} {\bibfnamefont {M.}~\bibnamefont {Wakihara}},\ }\href
  {\doibase https://doi.org/10.1006/jssc.1995.1001} {\bibfield  {journal}
  {\bibinfo  {journal} {Journal of Solid State Chemistry}\ }\textbf {\bibinfo
  {volume} {114}},\ \bibinfo {pages} {1} (\bibinfo {year} {1995})}\BibitemShut
  {NoStop}%
\bibitem [{\citenamefont {Onuki}\ \emph {et~al.}(1986)\citenamefont {Onuki},
  \citenamefont {Ina}, \citenamefont {Hirai},\ and\ \citenamefont
  {Komatsubara}}]{Onuki1986}%
  \BibitemOpen
  \bibfield  {author} {\bibinfo {author} {\bibfnamefont {Y.}~\bibnamefont
  {Onuki}}, \bibinfo {author} {\bibfnamefont {K.}~\bibnamefont {Ina}}, \bibinfo
  {author} {\bibfnamefont {T.}~\bibnamefont {Hirai}}, \ and\ \bibinfo {author}
  {\bibfnamefont {T.}~\bibnamefont {Komatsubara}},\ }\href {\doibase
  10.1143/JPSJ.55.347} {\bibfield  {journal} {\bibinfo  {journal} {Journal of
  the Physical Society of Japan}\ }\textbf {\bibinfo {volume} {55}},\ \bibinfo
  {pages} {347} (\bibinfo {year} {1986})}\BibitemShut {NoStop}%
\bibitem [{\citenamefont {Li}\ \emph {et~al.}(2011)\citenamefont {Li},
  \citenamefont {Lu}, \citenamefont {Zhu}, \citenamefont {Zhu}, \citenamefont
  {Yang}, \citenamefont {Song},\ and\ \citenamefont {Sun}}]{Li2011}%
  \BibitemOpen
  \bibfield  {author} {\bibinfo {author} {\bibfnamefont {L.}~\bibnamefont
  {Li}}, \bibinfo {author} {\bibfnamefont {W.}~\bibnamefont {Lu}}, \bibinfo
  {author} {\bibfnamefont {X.}~\bibnamefont {Zhu}}, \bibinfo {author}
  {\bibfnamefont {X.}~\bibnamefont {Zhu}}, \bibinfo {author} {\bibfnamefont
  {Z.}~\bibnamefont {Yang}}, \bibinfo {author} {\bibfnamefont {W.}~\bibnamefont
  {Song}}, \ and\ \bibinfo {author} {\bibfnamefont {Y.}~\bibnamefont {Sun}},\
  }\href {\doibase https://doi.org/10.1016/j.jmmm.2011.04.002} {\bibfield
  {journal} {\bibinfo  {journal} {Journal of Magnetism and Magnetic Materials}\
  }\textbf {\bibinfo {volume} {323}},\ \bibinfo {pages} {2536} (\bibinfo {year}
  {2011})}\BibitemShut {NoStop}%
\bibitem [{\citenamefont {Shand}\ \emph {et~al.}(2015)\citenamefont {Shand},
  \citenamefont {Cooling}, \citenamefont {Mellinger}, \citenamefont {Danker},
  \citenamefont {Kidd}, \citenamefont {Boyle},\ and\ \citenamefont
  {Strauss}}]{Shand2015}%
  \BibitemOpen
  \bibfield  {author} {\bibinfo {author} {\bibfnamefont {P.}~\bibnamefont
  {Shand}}, \bibinfo {author} {\bibfnamefont {C.}~\bibnamefont {Cooling}},
  \bibinfo {author} {\bibfnamefont {C.}~\bibnamefont {Mellinger}}, \bibinfo
  {author} {\bibfnamefont {J.}~\bibnamefont {Danker}}, \bibinfo {author}
  {\bibfnamefont {T.}~\bibnamefont {Kidd}}, \bibinfo {author} {\bibfnamefont
  {K.}~\bibnamefont {Boyle}}, \ and\ \bibinfo {author} {\bibfnamefont
  {L.}~\bibnamefont {Strauss}},\ }\href {\doibase
  https://doi.org/10.1016/j.jmmm.2015.01.023} {\bibfield  {journal} {\bibinfo
  {journal} {Journal of Magnetism and Magnetic Materials}\ }\textbf {\bibinfo
  {volume} {382}},\ \bibinfo {pages} {49} (\bibinfo {year} {2015})}\BibitemShut
  {NoStop}%
\bibitem [{\citenamefont {Zhang}\ \emph {et~al.}(2018)\citenamefont {Zhang},
  \citenamefont {Wei}, \citenamefont {Zheng}, \citenamefont {Lu}, \citenamefont
  {Wu}, \citenamefont {Zhu}, \citenamefont {Tang}, \citenamefont {Ning},
  \citenamefont {Han}, \citenamefont {Ling}, \citenamefont {Yang},
  \citenamefont {Gao}, \citenamefont {Qin},\ and\ \citenamefont
  {Tian}}]{Zhang2018}%
  \BibitemOpen
  \bibfield  {author} {\bibinfo {author} {\bibfnamefont {H.}~\bibnamefont
  {Zhang}}, \bibinfo {author} {\bibfnamefont {W.}~\bibnamefont {Wei}}, \bibinfo
  {author} {\bibfnamefont {G.}~\bibnamefont {Zheng}}, \bibinfo {author}
  {\bibfnamefont {J.}~\bibnamefont {Lu}}, \bibinfo {author} {\bibfnamefont
  {M.}~\bibnamefont {Wu}}, \bibinfo {author} {\bibfnamefont {X.}~\bibnamefont
  {Zhu}}, \bibinfo {author} {\bibfnamefont {J.}~\bibnamefont {Tang}}, \bibinfo
  {author} {\bibfnamefont {W.}~\bibnamefont {Ning}}, \bibinfo {author}
  {\bibfnamefont {Y.}~\bibnamefont {Han}}, \bibinfo {author} {\bibfnamefont
  {L.}~\bibnamefont {Ling}}, \bibinfo {author} {\bibfnamefont {J.}~\bibnamefont
  {Yang}}, \bibinfo {author} {\bibfnamefont {W.}~\bibnamefont {Gao}}, \bibinfo
  {author} {\bibfnamefont {Y.}~\bibnamefont {Qin}}, \ and\ \bibinfo {author}
  {\bibfnamefont {M.}~\bibnamefont {Tian}},\ }\href {\doibase
  10.1063/1.5034502} {\bibfield  {journal} {\bibinfo  {journal} {Applied
  Physics Letters}\ }\textbf {\bibinfo {volume} {113}},\ \bibinfo {pages}
  {072402} (\bibinfo {year} {2018})}\BibitemShut {NoStop}%
\bibitem [{\citenamefont {{Van Laar}}\ \emph {et~al.}(1971)\citenamefont {{Van
  Laar}}, \citenamefont {Rietveld},\ and\ \citenamefont {Ijdo}}]{VanLaar1971}%
  \BibitemOpen
  \bibfield  {author} {\bibinfo {author} {\bibfnamefont {B.}~\bibnamefont {{Van
  Laar}}}, \bibinfo {author} {\bibfnamefont {H.}~\bibnamefont {Rietveld}}, \
  and\ \bibinfo {author} {\bibfnamefont {D.}~\bibnamefont {Ijdo}},\ }\href
  {\doibase https://doi.org/10.1016/0022-4596(71)90019-3} {\bibfield  {journal}
  {\bibinfo  {journal} {Journal of Solid State Chemistry}\ }\textbf {\bibinfo
  {volume} {3}},\ \bibinfo {pages} {154} (\bibinfo {year} {1971})}\BibitemShut
  {NoStop}%
\bibitem [{\citenamefont {Dressel}\ and\ \citenamefont
  {Gruner}(2002)}]{Dressel2002}%
  \BibitemOpen
  \bibfield  {author} {\bibinfo {author} {\bibfnamefont {M.}~\bibnamefont
  {Dressel}}\ and\ \bibinfo {author} {\bibfnamefont {G.}~\bibnamefont
  {Gruner}},\ }\href {\doibase 10.1017/CBO9780511606168} {\emph {\bibinfo
  {title} {{Electrodynamics of Solids}}}}\ (\bibinfo  {publisher} {Cambridge
  University Press},\ \bibinfo {address} {Cambridge},\ \bibinfo {year}
  {2002})\BibitemShut {NoStop}%
\bibitem [{\citenamefont {Bari\ifmmode \check{s}\else
  \v{s}\fi{}i\ifmmode~\acute{c}\else \'{c}\fi{}}\ \emph
  {et~al.}(2011)\citenamefont {Bari\ifmmode \check{s}\else
  \v{s}\fi{}i\ifmmode~\acute{c}\else \'{c}\fi{}}, \citenamefont
  {Smiljani\ifmmode~\acute{c}\else \'{c}\fi{}}, \citenamefont {Pop\ifmmode
  \check{c}\else \v{c}\fi{}evi\ifmmode~\acute{c}\else \'{c}\fi{}},
  \citenamefont {Bilu\ifmmode \check{s}\else \v{s}\fi{}i\ifmmode~\acute{c}\else
  \'{c}\fi{}}, \citenamefont {Tuti\ifmmode~\check{s}\else \v{s}\fi{}},
  \citenamefont {Smontara}, \citenamefont {Berger}, \citenamefont {Ja\ifmmode
  \acute{c}\else \'{c}\fi{}imovi\ifmmode~\acute{c}\else \'{c}\fi{}},
  \citenamefont {Yuli},\ and\ \citenamefont {Forr\'o}}]{Barisic2011}%
  \BibitemOpen
  \bibfield  {author} {\bibinfo {author} {\bibfnamefont {N.}~\bibnamefont
  {Bari\ifmmode \check{s}\else \v{s}\fi{}i\ifmmode~\acute{c}\else \'{c}\fi{}}},
  \bibinfo {author} {\bibfnamefont {I.}~\bibnamefont
  {Smiljani\ifmmode~\acute{c}\else \'{c}\fi{}}}, \bibinfo {author}
  {\bibfnamefont {P.}~\bibnamefont {Pop\ifmmode \check{c}\else
  \v{c}\fi{}evi\ifmmode~\acute{c}\else \'{c}\fi{}}}, \bibinfo {author}
  {\bibfnamefont {A.}~\bibnamefont {Bilu\ifmmode \check{s}\else
  \v{s}\fi{}i\ifmmode~\acute{c}\else \'{c}\fi{}}}, \bibinfo {author}
  {\bibfnamefont {E.}~\bibnamefont {Tuti\ifmmode~\check{s}\else \v{s}\fi{}}},
  \bibinfo {author} {\bibfnamefont {A.}~\bibnamefont {Smontara}}, \bibinfo
  {author} {\bibfnamefont {H.}~\bibnamefont {Berger}}, \bibinfo {author}
  {\bibfnamefont {J.}~\bibnamefont {Ja\ifmmode \acute{c}\else
  \'{c}\fi{}imovi\ifmmode~\acute{c}\else \'{c}\fi{}}}, \bibinfo {author}
  {\bibfnamefont {O.}~\bibnamefont {Yuli}}, \ and\ \bibinfo {author}
  {\bibfnamefont {L.}~\bibnamefont {Forr\'o}},\ }\href {\doibase
  10.1103/PhysRevB.84.075157} {\bibfield  {journal} {\bibinfo  {journal} {Phys.
  Rev. B}\ }\textbf {\bibinfo {volume} {84}},\ \bibinfo {pages} {075157}
  (\bibinfo {year} {2011})}\BibitemShut {NoStop}%
\bibitem [{\citenamefont {Polesya}\ \emph {et~al.}(2019)\citenamefont
  {Polesya}, \citenamefont {Mankovsky},\ and\ \citenamefont
  {Ebert}}]{Polesya2019}%
  \BibitemOpen
  \bibfield  {author} {\bibinfo {author} {\bibfnamefont {S.}~\bibnamefont
  {Polesya}}, \bibinfo {author} {\bibfnamefont {S.}~\bibnamefont {Mankovsky}},
  \ and\ \bibinfo {author} {\bibfnamefont {H.}~\bibnamefont {Ebert}},\ }\href
  {\doibase doi:10.1515/znb-2018-0173} {\bibfield  {journal} {\bibinfo
  {journal} {Zeitschrift f\"ur Naturforschung B}\ }\textbf {\bibinfo {volume}
  {74}},\ \bibinfo {pages} {91} (\bibinfo {year} {2019})}\BibitemShut {NoStop}%
\bibitem [{Str()}]{Structure}%
  \BibitemOpen
  \href@noop {} {}\bibinfo {note} {2H-TaS$_2$ and 2H-NbS$_2$ share the same
  $P6_322$ space group \cite{Liu2022,SM,Polesya2019}. Since Ta and Nb belong to
  the same chemical group, it is reasonable to expect that their electronic
  band structure will be equivalent, beyond the band filling.}\BibitemShut
  {Stop}%
\bibitem [{\citenamefont {Hulliger}\ and\ \citenamefont
  {Pobitschka}(1970)}]{Hulliger1970}%
  \BibitemOpen
  \bibfield  {author} {\bibinfo {author} {\bibfnamefont {F.}~\bibnamefont
  {Hulliger}}\ and\ \bibinfo {author} {\bibfnamefont {E.}~\bibnamefont
  {Pobitschka}},\ }\href {\doibase
  https://doi.org/10.1016/0022-4596(70)90001-0} {\bibfield  {journal} {\bibinfo
   {journal} {Journal of Solid State Chemistry}\ }\textbf {\bibinfo {volume}
  {1}},\ \bibinfo {pages} {117} (\bibinfo {year} {1970})}\BibitemShut {NoStop}%
\bibitem [{\citenamefont {Miyadai}\ \emph {et~al.}(1983)\citenamefont
  {Miyadai}, \citenamefont {Kikuchi}, \citenamefont {Kondo}, \citenamefont
  {Sakka}, \citenamefont {Arai},\ and\ \citenamefont {Ishikawa}}]{Miyadai1983}%
  \BibitemOpen
  \bibfield  {author} {\bibinfo {author} {\bibfnamefont {T.}~\bibnamefont
  {Miyadai}}, \bibinfo {author} {\bibfnamefont {K.}~\bibnamefont {Kikuchi}},
  \bibinfo {author} {\bibfnamefont {H.}~\bibnamefont {Kondo}}, \bibinfo
  {author} {\bibfnamefont {S.}~\bibnamefont {Sakka}}, \bibinfo {author}
  {\bibfnamefont {M.}~\bibnamefont {Arai}}, \ and\ \bibinfo {author}
  {\bibfnamefont {Y.}~\bibnamefont {Ishikawa}},\ }\href {\doibase
  10.1143/JPSJ.52.1394} {\bibfield  {journal} {\bibinfo  {journal} {Journal of
  the Physical Society of Japan}\ }\textbf {\bibinfo {volume} {52}},\ \bibinfo
  {pages} {1394} (\bibinfo {year} {1983})}\BibitemShut {NoStop}%
\bibitem [{\citenamefont {Kousaka}\ \emph {et~al.}(2009)\citenamefont
  {Kousaka}, \citenamefont {Nakao}, \citenamefont {Kishine}, \citenamefont
  {Akita}, \citenamefont {Inoue},\ and\ \citenamefont
  {Akimitsu}}]{Kousaka2009}%
  \BibitemOpen
  \bibfield  {author} {\bibinfo {author} {\bibfnamefont {Y.}~\bibnamefont
  {Kousaka}}, \bibinfo {author} {\bibfnamefont {Y.}~\bibnamefont {Nakao}},
  \bibinfo {author} {\bibfnamefont {J.}~\bibnamefont {Kishine}}, \bibinfo
  {author} {\bibfnamefont {M.}~\bibnamefont {Akita}}, \bibinfo {author}
  {\bibfnamefont {K.}~\bibnamefont {Inoue}}, \ and\ \bibinfo {author}
  {\bibfnamefont {J.}~\bibnamefont {Akimitsu}},\ }\href {\doibase
  https://doi.org/10.1016/j.nima.2008.11.040} {\bibfield  {journal} {\bibinfo
  {journal} {Nuclear Instruments and Methods in Physics Research Section A:
  Accelerators, Spectrometers, Detectors and Associated Equipment}\ }\textbf
  {\bibinfo {volume} {600}},\ \bibinfo {pages} {250} (\bibinfo {year}
  {2009})}\BibitemShut {NoStop}%
\bibitem [{\citenamefont {Gorochov}\ \emph {et~al.}(1981)\citenamefont
  {Gorochov}, \citenamefont {Blanc-soreau}, \citenamefont {Rouxel},
  \citenamefont {Imbert},\ and\ \citenamefont {Jehanno}}]{Gorochov1981}%
  \BibitemOpen
  \bibfield  {author} {\bibinfo {author} {\bibfnamefont {O.}~\bibnamefont
  {Gorochov}}, \bibinfo {author} {\bibfnamefont {A.~L.}\ \bibnamefont
  {Blanc-soreau}}, \bibinfo {author} {\bibfnamefont {J.}~\bibnamefont
  {Rouxel}}, \bibinfo {author} {\bibfnamefont {P.}~\bibnamefont {Imbert}}, \
  and\ \bibinfo {author} {\bibfnamefont {G.}~\bibnamefont {Jehanno}},\ }\href
  {\doibase 10.1080/01418638108222164} {\bibfield  {journal} {\bibinfo
  {journal} {Philosophical Magazine B}\ }\textbf {\bibinfo {volume} {43}},\
  \bibinfo {pages} {621} (\bibinfo {year} {1981})}\BibitemShut {NoStop}%
\bibitem [{\citenamefont {Yamamura}\ \emph {et~al.}(2004)\citenamefont
  {Yamamura}, \citenamefont {Moriyama}, \citenamefont {Tsuji}, \citenamefont
  {Iwasa}, \citenamefont {Koyano}, \citenamefont {Katayama},\ and\
  \citenamefont {Ito}}]{Yamamura2004}%
  \BibitemOpen
  \bibfield  {author} {\bibinfo {author} {\bibfnamefont {Y.}~\bibnamefont
  {Yamamura}}, \bibinfo {author} {\bibfnamefont {S.}~\bibnamefont {Moriyama}},
  \bibinfo {author} {\bibfnamefont {T.}~\bibnamefont {Tsuji}}, \bibinfo
  {author} {\bibfnamefont {Y.}~\bibnamefont {Iwasa}}, \bibinfo {author}
  {\bibfnamefont {M.}~\bibnamefont {Koyano}}, \bibinfo {author} {\bibfnamefont
  {S.}~\bibnamefont {Katayama}}, \ and\ \bibinfo {author} {\bibfnamefont
  {M.}~\bibnamefont {Ito}},\ }\href {\doibase
  https://doi.org/10.1016/j.jallcom.2004.04.045} {\bibfield  {journal}
  {\bibinfo  {journal} {Journal of Alloys and Compounds}\ }\textbf {\bibinfo
  {volume} {383}},\ \bibinfo {pages} {338} (\bibinfo {year} {2004})},\ \bibinfo
  {note} {proceedings of the 14th International Conference on Solid Compounds
  of Transition Elements (SCTE 2003)}\BibitemShut {NoStop}%
\bibitem [{\citenamefont {Mankovsky}\ \emph {et~al.}(2016)\citenamefont
  {Mankovsky}, \citenamefont {Polesya}, \citenamefont {Ebert},\ and\
  \citenamefont {Bensch}}]{Mankovsky2016}%
  \BibitemOpen
  \bibfield  {author} {\bibinfo {author} {\bibfnamefont {S.}~\bibnamefont
  {Mankovsky}}, \bibinfo {author} {\bibfnamefont {S.}~\bibnamefont {Polesya}},
  \bibinfo {author} {\bibfnamefont {H.}~\bibnamefont {Ebert}}, \ and\ \bibinfo
  {author} {\bibfnamefont {W.}~\bibnamefont {Bensch}},\ }\href {\doibase
  10.1103/PhysRevB.94.184430} {\bibfield  {journal} {\bibinfo  {journal} {Phys.
  Rev. B}\ }\textbf {\bibinfo {volume} {94}},\ \bibinfo {pages} {184430}
  (\bibinfo {year} {2016})}\BibitemShut {NoStop}%
\bibitem [{\citenamefont {Polesya}\ \emph {et~al.}(2020)\citenamefont
  {Polesya}, \citenamefont {Mankovsky}, \citenamefont {Ebert}, \citenamefont
  {Naumov}, \citenamefont {ElGhazali}, \citenamefont {Schnelle}, \citenamefont
  {Medvedev}, \citenamefont {Mangelsen},\ and\ \citenamefont
  {Bensch}}]{Polesya2020}%
  \BibitemOpen
  \bibfield  {author} {\bibinfo {author} {\bibfnamefont {S.}~\bibnamefont
  {Polesya}}, \bibinfo {author} {\bibfnamefont {S.}~\bibnamefont {Mankovsky}},
  \bibinfo {author} {\bibfnamefont {H.}~\bibnamefont {Ebert}}, \bibinfo
  {author} {\bibfnamefont {P.~G.}\ \bibnamefont {Naumov}}, \bibinfo {author}
  {\bibfnamefont {M.~A.}\ \bibnamefont {ElGhazali}}, \bibinfo {author}
  {\bibfnamefont {W.}~\bibnamefont {Schnelle}}, \bibinfo {author}
  {\bibfnamefont {S.}~\bibnamefont {Medvedev}}, \bibinfo {author}
  {\bibfnamefont {S.}~\bibnamefont {Mangelsen}}, \ and\ \bibinfo {author}
  {\bibfnamefont {W.}~\bibnamefont {Bensch}},\ }\href {\doibase
  10.1103/PhysRevB.102.174423} {\bibfield  {journal} {\bibinfo  {journal}
  {Phys. Rev. B}\ }\textbf {\bibinfo {volume} {102}},\ \bibinfo {pages}
  {174423} (\bibinfo {year} {2020})}\BibitemShut {NoStop}%
\bibitem [{\citenamefont {Hu}\ \emph {et~al.}(2008)\citenamefont {Hu},
  \citenamefont {Wang}, \citenamefont {Hu}, \citenamefont {Petrovic},
  \citenamefont {Morosan}, \citenamefont {Cava}, \citenamefont {Fang},\ and\
  \citenamefont {Wang}}]{Hu2008}%
  \BibitemOpen
  \bibfield  {author} {\bibinfo {author} {\bibfnamefont {W.~Z.}\ \bibnamefont
  {Hu}}, \bibinfo {author} {\bibfnamefont {G.~T.}\ \bibnamefont {Wang}},
  \bibinfo {author} {\bibfnamefont {R.}~\bibnamefont {Hu}}, \bibinfo {author}
  {\bibfnamefont {C.}~\bibnamefont {Petrovic}}, \bibinfo {author}
  {\bibfnamefont {E.}~\bibnamefont {Morosan}}, \bibinfo {author} {\bibfnamefont
  {R.~J.}\ \bibnamefont {Cava}}, \bibinfo {author} {\bibfnamefont
  {Z.}~\bibnamefont {Fang}}, \ and\ \bibinfo {author} {\bibfnamefont {N.~L.}\
  \bibnamefont {Wang}},\ }\href {\doibase 10.1103/PhysRevB.78.085120}
  {\bibfield  {journal} {\bibinfo  {journal} {Phys. Rev. B}\ }\textbf {\bibinfo
  {volume} {78}},\ \bibinfo {pages} {085120} (\bibinfo {year}
  {2008})}\BibitemShut {NoStop}%
\bibitem [{\citenamefont {Li}\ \emph {et~al.}(2020)\citenamefont {Li},
  \citenamefont {Zhang}, \citenamefont {Myeong}, \citenamefont {Shin},
  \citenamefont {Lim}, \citenamefont {Kim}, \citenamefont {Kim}, \citenamefont
  {Jin}, \citenamefont {Cavill}, \citenamefont {Kim}, \citenamefont {Kim},
  \citenamefont {Lischner}, \citenamefont {Ferreira},\ and\ \citenamefont
  {Cho}}]{Li2020}%
  \BibitemOpen
  \bibfield  {author} {\bibinfo {author} {\bibfnamefont {L.}~\bibnamefont
  {Li}}, \bibinfo {author} {\bibfnamefont {J.}~\bibnamefont {Zhang}}, \bibinfo
  {author} {\bibfnamefont {G.}~\bibnamefont {Myeong}}, \bibinfo {author}
  {\bibfnamefont {W.}~\bibnamefont {Shin}}, \bibinfo {author} {\bibfnamefont
  {H.}~\bibnamefont {Lim}}, \bibinfo {author} {\bibfnamefont {B.}~\bibnamefont
  {Kim}}, \bibinfo {author} {\bibfnamefont {S.}~\bibnamefont {Kim}}, \bibinfo
  {author} {\bibfnamefont {T.}~\bibnamefont {Jin}}, \bibinfo {author}
  {\bibfnamefont {S.}~\bibnamefont {Cavill}}, \bibinfo {author} {\bibfnamefont
  {B.~S.}\ \bibnamefont {Kim}}, \bibinfo {author} {\bibfnamefont
  {C.}~\bibnamefont {Kim}}, \bibinfo {author} {\bibfnamefont {J.}~\bibnamefont
  {Lischner}}, \bibinfo {author} {\bibfnamefont {A.}~\bibnamefont {Ferreira}},
  \ and\ \bibinfo {author} {\bibfnamefont {S.}~\bibnamefont {Cho}},\ }\bibfield
   {booktitle} {\emph {\bibinfo {booktitle} {ACS Nano}},\ }\href {\doibase
  10.1021/acsnano.0c01037} {\bibfield  {journal} {\bibinfo  {journal} {ACS
  Nano}\ }\textbf {\bibinfo {volume} {14}},\ \bibinfo {pages} {5251} (\bibinfo
  {year} {2020})}\BibitemShut {NoStop}%
\bibitem [{\citenamefont {Xiang}\ \emph {et~al.}(2020)\citenamefont {Xiang},
  \citenamefont {Inoue}, \citenamefont {Zheng}, \citenamefont {Kumamoto},
  \citenamefont {Qian}, \citenamefont {Sato}, \citenamefont {Liu},
  \citenamefont {Tang}, \citenamefont {Gokhale}, \citenamefont {Guo},
  \citenamefont {Hisama}, \citenamefont {Yotsumoto}, \citenamefont {Ogamoto},
  \citenamefont {Arai}, \citenamefont {Kobayashi}, \citenamefont {Zhang},
  \citenamefont {Hou}, \citenamefont {Anisimov}, \citenamefont {Maruyama},
  \citenamefont {Miyata}, \citenamefont {Okada}, \citenamefont {Chiashi},
  \citenamefont {Li}, \citenamefont {Kong}, \citenamefont {Kauppinen},
  \citenamefont {Ikuhara}, \citenamefont {Suenaga},\ and\ \citenamefont
  {Maruyama}}]{Xiang2020}%
  \BibitemOpen
  \bibfield  {author} {\bibinfo {author} {\bibfnamefont {R.}~\bibnamefont
  {Xiang}}, \bibinfo {author} {\bibfnamefont {T.}~\bibnamefont {Inoue}},
  \bibinfo {author} {\bibfnamefont {Y.}~\bibnamefont {Zheng}}, \bibinfo
  {author} {\bibfnamefont {A.}~\bibnamefont {Kumamoto}}, \bibinfo {author}
  {\bibfnamefont {Y.}~\bibnamefont {Qian}}, \bibinfo {author} {\bibfnamefont
  {Y.}~\bibnamefont {Sato}}, \bibinfo {author} {\bibfnamefont {M.}~\bibnamefont
  {Liu}}, \bibinfo {author} {\bibfnamefont {D.}~\bibnamefont {Tang}}, \bibinfo
  {author} {\bibfnamefont {D.}~\bibnamefont {Gokhale}}, \bibinfo {author}
  {\bibfnamefont {J.}~\bibnamefont {Guo}}, \bibinfo {author} {\bibfnamefont
  {K.}~\bibnamefont {Hisama}}, \bibinfo {author} {\bibfnamefont
  {S.}~\bibnamefont {Yotsumoto}}, \bibinfo {author} {\bibfnamefont
  {T.}~\bibnamefont {Ogamoto}}, \bibinfo {author} {\bibfnamefont
  {H.}~\bibnamefont {Arai}}, \bibinfo {author} {\bibfnamefont {Y.}~\bibnamefont
  {Kobayashi}}, \bibinfo {author} {\bibfnamefont {H.}~\bibnamefont {Zhang}},
  \bibinfo {author} {\bibfnamefont {B.}~\bibnamefont {Hou}}, \bibinfo {author}
  {\bibfnamefont {A.}~\bibnamefont {Anisimov}}, \bibinfo {author}
  {\bibfnamefont {M.}~\bibnamefont {Maruyama}}, \bibinfo {author}
  {\bibfnamefont {Y.}~\bibnamefont {Miyata}}, \bibinfo {author} {\bibfnamefont
  {S.}~\bibnamefont {Okada}}, \bibinfo {author} {\bibfnamefont
  {S.}~\bibnamefont {Chiashi}}, \bibinfo {author} {\bibfnamefont
  {Y.}~\bibnamefont {Li}}, \bibinfo {author} {\bibfnamefont {J.}~\bibnamefont
  {Kong}}, \bibinfo {author} {\bibfnamefont {E.~I.}\ \bibnamefont {Kauppinen}},
  \bibinfo {author} {\bibfnamefont {Y.}~\bibnamefont {Ikuhara}}, \bibinfo
  {author} {\bibfnamefont {K.}~\bibnamefont {Suenaga}}, \ and\ \bibinfo
  {author} {\bibfnamefont {S.}~\bibnamefont {Maruyama}},\ }\href {\doibase
  10.1126/science.aaz2570} {\bibfield  {journal} {\bibinfo  {journal}
  {Science}\ }\textbf {\bibinfo {volume} {367}},\ \bibinfo {pages} {537}
  (\bibinfo {year} {2020})}\BibitemShut {NoStop}%
\bibitem [{\citenamefont {Li}\ \emph {et~al.}(2010)\citenamefont {Li},
  \citenamefont {Stender}, \citenamefont {Ringe}, \citenamefont {Marks},\ and\
  \citenamefont {Odom}}]{Li2010}%
  \BibitemOpen
  \bibfield  {author} {\bibinfo {author} {\bibfnamefont {P.}~\bibnamefont
  {Li}}, \bibinfo {author} {\bibfnamefont {C.~L.}\ \bibnamefont {Stender}},
  \bibinfo {author} {\bibfnamefont {E.}~\bibnamefont {Ringe}}, \bibinfo
  {author} {\bibfnamefont {L.~D.}\ \bibnamefont {Marks}}, \ and\ \bibinfo
  {author} {\bibfnamefont {T.~W.}\ \bibnamefont {Odom}},\ }\href {\doibase
  https://doi.org/10.1002/smll.201000226} {\bibfield  {journal} {\bibinfo
  {journal} {Small}\ }\textbf {\bibinfo {volume} {6}},\ \bibinfo {pages} {1096}
  (\bibinfo {year} {2010})}\BibitemShut {NoStop}%
\bibitem [{\citenamefont {Zhang}\ \emph {et~al.}(2022)\citenamefont {Zhang},
  \citenamefont {Algaidi}, \citenamefont {Li}, \citenamefont {Yuan},\ and\
  \citenamefont {Zhang}}]{Zhang2022}%
  \BibitemOpen
  \bibfield  {author} {\bibinfo {author} {\bibfnamefont {C.-H.}\ \bibnamefont
  {Zhang}}, \bibinfo {author} {\bibfnamefont {H.}~\bibnamefont {Algaidi}},
  \bibinfo {author} {\bibfnamefont {P.}~\bibnamefont {Li}}, \bibinfo {author}
  {\bibfnamefont {Y.}~\bibnamefont {Yuan}}, \ and\ \bibinfo {author}
  {\bibfnamefont {X.-X.}\ \bibnamefont {Zhang}},\ }\href {\doibase
  10.1007/s12598-022-02037-7} {\bibfield  {journal} {\bibinfo  {journal} {Rare
  Metals}\ }\textbf {\bibinfo {volume} {41}},\ \bibinfo {pages} {3005}
  (\bibinfo {year} {2022})}\BibitemShut {NoStop}%
\bibitem [{\citenamefont {Zheng}\ \emph {et~al.}(2021)\citenamefont {Zheng},
  \citenamefont {Wang}, \citenamefont {Zhu}, \citenamefont {Tan}, \citenamefont
  {Wang}, \citenamefont {Albarakati}, \citenamefont {Aloufi}, \citenamefont
  {Algarni}, \citenamefont {Farrar}, \citenamefont {Wu}, \citenamefont {Yao},
  \citenamefont {Tian}, \citenamefont {Zhou},\ and\ \citenamefont
  {Wang}}]{Zheng2021}%
  \BibitemOpen
  \bibfield  {author} {\bibinfo {author} {\bibfnamefont {G.}~\bibnamefont
  {Zheng}}, \bibinfo {author} {\bibfnamefont {M.}~\bibnamefont {Wang}},
  \bibinfo {author} {\bibfnamefont {X.}~\bibnamefont {Zhu}}, \bibinfo {author}
  {\bibfnamefont {C.}~\bibnamefont {Tan}}, \bibinfo {author} {\bibfnamefont
  {J.}~\bibnamefont {Wang}}, \bibinfo {author} {\bibfnamefont {S.}~\bibnamefont
  {Albarakati}}, \bibinfo {author} {\bibfnamefont {N.}~\bibnamefont {Aloufi}},
  \bibinfo {author} {\bibfnamefont {M.}~\bibnamefont {Algarni}}, \bibinfo
  {author} {\bibfnamefont {L.}~\bibnamefont {Farrar}}, \bibinfo {author}
  {\bibfnamefont {M.}~\bibnamefont {Wu}}, \bibinfo {author} {\bibfnamefont
  {Y.}~\bibnamefont {Yao}}, \bibinfo {author} {\bibfnamefont {M.}~\bibnamefont
  {Tian}}, \bibinfo {author} {\bibfnamefont {J.}~\bibnamefont {Zhou}}, \ and\
  \bibinfo {author} {\bibfnamefont {L.}~\bibnamefont {Wang}},\ }\href {\doibase
  10.1038/s41467-021-23658-z} {\bibfield  {journal} {\bibinfo  {journal}
  {Nature Communications}\ }\textbf {\bibinfo {volume} {12}},\ \bibinfo {pages}
  {3639} (\bibinfo {year} {2021})}\BibitemShut {NoStop}%
\bibitem [{\citenamefont {Hu}\ \emph {et~al.}(2007)\citenamefont {Hu},
  \citenamefont {Li}, \citenamefont {Yan}, \citenamefont {Wen}, \citenamefont
  {Wu}, \citenamefont {Chen},\ and\ \citenamefont {Wang}}]{Hu2007}%
  \BibitemOpen
  \bibfield  {author} {\bibinfo {author} {\bibfnamefont {W.~Z.}\ \bibnamefont
  {Hu}}, \bibinfo {author} {\bibfnamefont {G.}~\bibnamefont {Li}}, \bibinfo
  {author} {\bibfnamefont {J.}~\bibnamefont {Yan}}, \bibinfo {author}
  {\bibfnamefont {H.~H.}\ \bibnamefont {Wen}}, \bibinfo {author} {\bibfnamefont
  {G.}~\bibnamefont {Wu}}, \bibinfo {author} {\bibfnamefont {X.~H.}\
  \bibnamefont {Chen}}, \ and\ \bibinfo {author} {\bibfnamefont {N.~L.}\
  \bibnamefont {Wang}},\ }\href {\doibase 10.1103/PhysRevB.76.045103}
  {\bibfield  {journal} {\bibinfo  {journal} {Phys. Rev. B}\ }\textbf {\bibinfo
  {volume} {76}},\ \bibinfo {pages} {045103} (\bibinfo {year}
  {2007})}\BibitemShut {NoStop}%
\bibitem [{\citenamefont {Maulana}\ \emph {et~al.}(2021)\citenamefont
  {Maulana}, \citenamefont {Li}, \citenamefont {Uykur}, \citenamefont {Manna},
  \citenamefont {Polatkan}, \citenamefont {Felser}, \citenamefont {Dressel},\
  and\ \citenamefont {Pronin}}]{Maulana2021}%
  \BibitemOpen
  \bibfield  {author} {\bibinfo {author} {\bibfnamefont {L.~Z.}\ \bibnamefont
  {Maulana}}, \bibinfo {author} {\bibfnamefont {Z.}~\bibnamefont {Li}},
  \bibinfo {author} {\bibfnamefont {E.}~\bibnamefont {Uykur}}, \bibinfo
  {author} {\bibfnamefont {K.}~\bibnamefont {Manna}}, \bibinfo {author}
  {\bibfnamefont {S.}~\bibnamefont {Polatkan}}, \bibinfo {author}
  {\bibfnamefont {C.}~\bibnamefont {Felser}}, \bibinfo {author} {\bibfnamefont
  {M.}~\bibnamefont {Dressel}}, \ and\ \bibinfo {author} {\bibfnamefont
  {A.~V.}\ \bibnamefont {Pronin}},\ }\href {\doibase
  10.1103/PhysRevB.103.115206} {\bibfield  {journal} {\bibinfo  {journal}
  {Phys. Rev. B}\ }\textbf {\bibinfo {volume} {103}},\ \bibinfo {pages}
  {115206} (\bibinfo {year} {2021})}\BibitemShut {NoStop}%
\bibitem [{\citenamefont {Corasaniti}\ \emph {et~al.}(2020)\citenamefont
  {Corasaniti}, \citenamefont {Yang}, \citenamefont {Sen}, \citenamefont
  {Willa}, \citenamefont {Merz}, \citenamefont {Haghighirad}, \citenamefont
  {Le~Tacon},\ and\ \citenamefont {Degiorgi}}]{Corasaniti2020}%
  \BibitemOpen
  \bibfield  {author} {\bibinfo {author} {\bibfnamefont {M.}~\bibnamefont
  {Corasaniti}}, \bibinfo {author} {\bibfnamefont {R.}~\bibnamefont {Yang}},
  \bibinfo {author} {\bibfnamefont {K.}~\bibnamefont {Sen}}, \bibinfo {author}
  {\bibfnamefont {K.}~\bibnamefont {Willa}}, \bibinfo {author} {\bibfnamefont
  {M.}~\bibnamefont {Merz}}, \bibinfo {author} {\bibfnamefont {A.~A.}\
  \bibnamefont {Haghighirad}}, \bibinfo {author} {\bibfnamefont
  {M.}~\bibnamefont {Le~Tacon}}, \ and\ \bibinfo {author} {\bibfnamefont
  {L.}~\bibnamefont {Degiorgi}},\ }\href {\doibase 10.1103/PhysRevB.102.161109}
  {\bibfield  {journal} {\bibinfo  {journal} {Phys. Rev. B}\ }\textbf {\bibinfo
  {volume} {102}},\ \bibinfo {pages} {161109} (\bibinfo {year}
  {2020})}\BibitemShut {NoStop}%
\end{thebibliography}

%

\end{document}